\documentclass[11pt]{article}

\PassOptionsToPackage{dvipsnames,table}{xcolor}
\usepackage[preprint]{acl}

\usepackage{times}
\usepackage{latexsym}

\usepackage[T1]{fontenc}
\usepackage[utf8]{inputenc}

\usepackage{microtype}

\usepackage{inconsolata}

\usepackage{graphicx}
\usepackage{url}
\usepackage{booktabs}
\usepackage{pifont}
\usepackage{graphicx}
\usepackage{wrapfig}
\usepackage{makecell}
\usepackage{subcaption}
\usepackage{amsmath}
\usepackage{algorithm}
\usepackage{algpseudocode}

\usepackage[dvipsnames,table]{xcolor}
\usepackage{colortbl}
\usepackage[most]{tcolorbox}
\usepackage{listings}

\usepackage{hyperref}
\PassOptionsToPackage{hyphens}{url}
\definecolor{darkblue}{rgb}{0, 0, 0.5}
\hypersetup{colorlinks=true, citecolor=darkblue, linkcolor=darkblue, urlcolor=darkblue}
\usepackage{xurl}

\definecolor{tableheader}{RGB}{182,216,210}
\definecolor{tablerow}{RGB}{220,239,235}
\definecolor{tablebest}{RGB}{195,225,219}

\definecolor{pastelblue}{RGB}{185,195,235}
\definecolor{pastelbluebg}{RGB}{232,236,250}
\definecolor{pastelgreen}{RGB}{175,220,200}
\definecolor{pastelgreenbg}{RGB}{228,245,236}
\definecolor{pastelrose}{RGB}{230,185,185}
\definecolor{pastelrosebg}{RGB}{248,232,232}
\definecolor{promptgray}{RGB}{245,245,248}
\definecolor{promptframe}{RGB}{200,200,210}

\newtcolorbox{systempromptbox}[1][]{%
  enhanced, breakable,
  colback=pastelbluebg, colframe=pastelblue, coltitle=black,
  fonttitle=\bfseries\small, title={#1},
  boxrule=0.6pt, arc=2pt,
  left=4pt, right=4pt, top=2pt, bottom=2pt,
  toptitle=2pt, bottomtitle=2pt,
  attach boxed title to top left={yshift=-2mm, xshift=4mm},
  boxed title style={colback=pastelblue, colframe=pastelblue, arc=1.5pt, boxrule=0pt},
}
\newtcolorbox{userpromptbox}[1][]{%
  enhanced, breakable,
  colback=pastelgreenbg, colframe=pastelgreen, coltitle=black,
  fonttitle=\bfseries\small, title={#1},
  boxrule=0.6pt, arc=2pt,
  left=4pt, right=4pt, top=2pt, bottom=2pt,
  toptitle=2pt, bottomtitle=2pt,
  attach boxed title to top left={yshift=-2mm, xshift=4mm},
  boxed title style={colback=pastelgreen, colframe=pastelgreen, arc=1.5pt, boxrule=0pt},
}
\newtcolorbox{retrypromptbox}[1][]{%
  enhanced, breakable,
  colback=pastelrosebg, colframe=pastelrose, coltitle=black,
  fonttitle=\bfseries\small, title={#1},
  boxrule=0.6pt, arc=2pt,
  left=4pt, right=4pt, top=2pt, bottom=2pt,
  toptitle=2pt, bottomtitle=2pt,
  attach boxed title to top left={yshift=-2mm, xshift=4mm},
  boxed title style={colback=pastelrose, colframe=pastelrose, arc=1.5pt, boxrule=0pt},
}

\lstdefinestyle{promptstyle}{
  basicstyle=\ttfamily\scriptsize,
  breaklines=true, breakatwhitespace=false,
  columns=fullflexible, keepspaces=true,
  frame=none, xleftmargin=0pt,
  aboveskip=0pt, belowskip=0pt,
}

\newcommand{\tool}{\textsc{CTI2TTP}}

\title{{\Large \tool: Evidence-Grounded TTP Extraction from Cyber Threat Intelligence via Diverge-then-Converge Verification}}

\author{Yutong Cheng \\
  Virginia Tech \\
  \And
  Changze Li \\
  Virginia Tech \\
  \And
  Raihan Sultan Pasha Basuki \\
  Virginia Tech \\
  \AND
  Qian Cui \\
  Amazon \\
  \And
  Wei Ding \\
  Amazon \\
  \And
  Peng Gao \\
  Virginia Tech}

\begin{document}
\maketitle
\begin{abstract}
\looseness=-1Extracting MITRE ATT\&CK techniques from cyber threat intelligence (CTI) reports is an open-set, multi-label problem requiring both high recall (not missing techniques) and high precision (not hallucinating unsupported ones). Existing rule-based, supervised, and LLM-based methods struggle to achieve both: rule-based and supervised approaches lack generalizability across diverse attack descriptions, while LLM-based approaches that couple candidate generation and validation within a single inference step suffer from limited recall and precision simultaneously. We propose \tool, which addresses this challenge through a diverge-then-converge design inspired by how human analysts work: first extracting broadly, then verifying rigorously. Reports are first decomposed into atomic behaviors, each anchored by a deterministic span localization stage to a specific evidence window in the source text. Candidate techniques are then proposed broadly for each behavior, and a convergent verification stage retains only candidates supported by both the localized evidence and the authoritative MITRE definition. We contribute two evaluation resources, a cleaned TRAM benchmark (TRAM-Clean) and a new annotated dataset (\tool-Bench), to address known annotation noise in existing benchmarks and elevate the task to document-level TTP extraction. On TRAM-Clean and \tool-Bench, \tool~achieves 76.48\% and 87.39\% macro-F1 respectively, outperforming chain-of-thought prompting with the same backbone by 63.5\% and 29.4\%.
\end{abstract}

\section{Introduction}
\label{sec:intro}

\begin{figure*}[t]
\centering
\includegraphics[width=\textwidth]{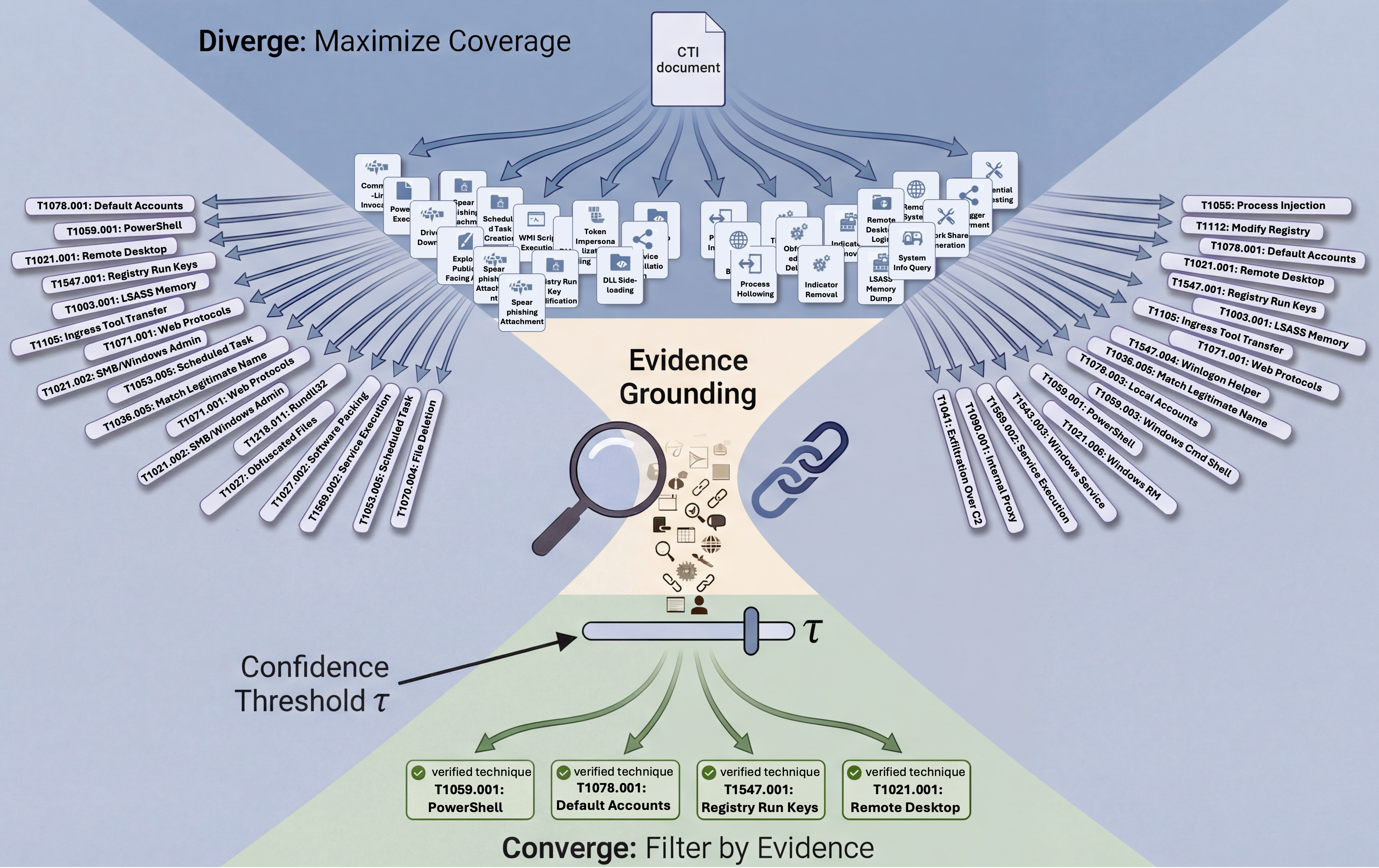}
\caption{The diverge-then-converge principle behind \tool. Existing single-pass methods couple precision and recall within one inference step. \tool~explicitly separates them: the \textbf{divergent phase} (top) decomposes a CTI report into atomic behaviors and proposes a broad set of candidate techniques, maximizing coverage. \textbf{Evidence grounding} (center) anchors each behavior to a localized sentence window in the original report, restricting downstream reasoning to verifiable textual support. The \textbf{convergent phase} (bottom) filters candidates through a verification threshold~$\tau$, retaining only techniques that are explicitly supported by the grounded evidence.}
\label{fig:overview}
\end{figure*}

Indicators of Compromise (IoCs) such as malicious IP addresses, file hashes, and domain names have long served as the primary currency of cyber threat intelligence (CTI). However, IoCs are inherently ephemeral: adversaries can rotate infrastructure, recompile binaries, and register new domains with minimal effort, rendering specific indicators obsolete within days~\citep{buechel2025sok, mitretrieval}. In contrast, tactics, techniques, and procedures (TTPs) describe \emph{how} attackers achieve their objectives, such as gaining initial access, moving laterally, escalating privileges, and exfiltrating data, and remain operationally valuable long after individual IoCs have expired~\citep{ttpdrill}. The MITRE ATT\&CK framework provides a standardized ontology for cataloging these TTPs. Mapping report content onto this ontology turns narrative descriptions of adversary behavior into precisely defined technique identifiers that carry the same meaning across reports, analysts, and tools. Automatically extracting TTPs from unstructured CTI reports has therefore become a central task in threat intelligence analysis~\citep{strom2020mitre}.

However, these reports embed multiple attack techniques in narrative descriptions of adversary behavior across dozens of paragraphs, often without explicit mention of MITRE identifiers. TTPs must be inferred from context rather than matched by keywords~\citep{ttpdrill, buechel2025sok}. As a result, TTP extraction is an open-set, multi-label problem. The core challenge is to simultaneously achieve high \emph{recall} (not missing techniques that are genuinely described) and high \emph{precision} (not hallucinating techniques that lack textual support).

\looseness=-1Existing methods fall into three paradigms. Rule-based systems such as TTPDrill~\citep{ttpdrill} offer interpretability but depend on manually constructed ontologies that cannot keep pace with evolving attack descriptions. Supervised neural models, including LADDER~\citep{ladder} and MITREtrieval~\citep{mitretrieval}, learn contextual patterns but require substantial annotation effort and degrade under distribution shift to new report styles or technique vocabularies~\citep{buechel2025sok}. Large language models (LLMs) have recently been applied to CTI extraction and analysis. CTINexus~\citep{cheng2025ctinexusautomaticcyberthreat} and AttacKG+~\citep{attackgplus} construct CTI knowledge graphs from reports, with AttacKG+ producing TTP labels as an intermediate product. However, no existing work applies LLMs to TTP extraction as a standalone task. High-recall, precise TTP extraction directly informs detection rule authoring, threat actor profiling, and defensive prioritization. Missed techniques leave blind spots in defensive coverage, while hallucinated ones waste analyst effort on nonexistent threats.

TTP extraction is inherently a bi-objective optimization problem of maximizing coverage over all techniques described in a report while precisely mapping each behavior to the correct MITRE ATT\&CK identifier. Under single-pass prompting, these two objectives are tightly coupled, introducing two compounding failure modes. First, the model may sacrifice coverage for confidence, omitting behaviors that are genuinely present but less salient within a dense, multi-technique passage. Second, even when a behavior is correctly extracted, a \emph{surface similarity mismatch}, a gap between the operational language used in CTI reports and the canonical terminology in MITRE ATT\&CK definitions, degrades the accuracy of the behavior-to-identifier mapping. For example, a report stating ``the malware writes itself to the \texttt{Run} registry key to survive reboots'' describes T1547 (Boot or Logon Autostart Execution),\footnote{\url{https://attack.mitre.org/techniques/T1547/}} yet the phrasing shares almost no lexical overlap with the technique name or its official definition, causing the model to select a superficially closer but incorrect technique.

We propose a \emph{diverge-then-converge} principle that \emph{separates broad candidate generation from verification} against localized evidence, and instantiate this principle in \tool, a four-stage pipeline for TTP extraction to address these challenges.
In the divergent phase, \tool~decomposes a CTI report into atomic attack behaviors and proposes a broad set of candidate MITRE ATT\&CK techniques for each behavior, deliberately favoring coverage over confidence. A deterministic \emph{span localization} stage anchors each behavior to a specific sentence window in the original report, restricting all downstream reasoning to verifiable textual evidence rather than the full document. In the convergent phase, a verification stage evaluates each candidate against both its localized evidence span and the official MITRE ATT\&CK technique description, retaining only those that exceed a confidence threshold~$\tau$. This design decouples the two failure modes. The divergent phase eliminates omission risk by treating extraction as a standalone objective, while the convergent phase resolves surface similarity mismatch by cross-referencing evidence against authoritative definitions. The threshold~$\tau$ gives practitioners direct, interpretable control over the precision–recall trade-off, and every final prediction is traceable to its source passage.

\looseness=-1To comprehensively benchmark \tool's performance in TTP extraction, we contribute two evaluation resources. First, we systematically clean the most widely used benchmark, TRAM~\citep{tram_dataset}, a sentence-level TTP evaluation dataset containing substantial annotation errors. The corrected version, TRAM-Clean, provides a reliable baseline for comparison with prior work. However, sentence-level evaluation alone cannot assess a system's ability to extract TTPs from complete reports, where techniques are distributed across dozens of paragraphs and must be identified without pre-segmentation. We therefore construct \tool-Bench, a document-level benchmark of 150 CTI reports that requires extracting TTPs directly from documents. Both datasets will be released to support further research.

\section{Related Work}

\paragraph{TTP Extraction from CTI Reports.}
Early approaches such as TTPDrill~\citep{ttpdrill} rely on rule-based matching using dependency parsing and predefined ontologies. These methods are easy to interpret but struggle when the same technique is described in different ways or only implied in the text. Later work moves toward supervised neural models. Systems such as EXTRACTOR~\citep{extractor}, AttacKG~\citep{attackkg}, LADDER~\citep{ladder}, and MITREtrieval~\citep{mitretrieval} typically treat the problem as classification or alignment over ATT\&CK techniques using fine-tuned transformer models. While they capture context better than rule-based methods, they depend on labeled data and often do not generalize well across different report styles~\citep{buechel2025sok}.
More recently, LLMs have been applied to CTI analysis tasks. AttacKG+~\citep{attackgplus} and CTINexus~\citep{cheng2025ctinexusautomaticcyberthreat} focus on CTI knowledge graph construction, where AttacKG+ treats TTP labeling as one layer of a multi-layered output that also includes behavior graphs and state summaries, mapping pre-extracted structured behavior graphs to ATT\&CK labels. In contrast, \tool~targets standalone TTP extraction from raw CTI reports.

\paragraph{LLM-Based Structured Extraction and Verification.}
Similar challenges appear in other domains such as medical coding, where long documents must be mapped to large label spaces. Recent approaches break the task into steps such as evidence extraction and candidate ranking~\citep{baksi-etal-2025-medcoder, clh_icd}, suggesting that a single LLM call is often insufficient. More broadly, LLM-based information extraction has been explored for relation extraction and document-level extraction tasks~\citep{swarup-etal-2025-llm4re, bhattacharyya-etal-2025-information}, as well as multi-stage pipelines that incorporate validation mechanisms for structured data extraction~\citep{aggarwal2025fiscalie}. To reduce hallucinations, methods such as Chain-of-Verification~\citep{dhuliawala-etal-2024-chain} and Self-Refine~\citep{madaan2023selfrefine} introduce additional reasoning or refinement steps. These approaches operate after generation and rely on the model to correct itself, which can be unreliable without external grounding~\citep{huang2024selfcorrection}. In contrast, \tool~restricts the input by localizing evidence spans with deterministic algorithms and verifies candidates against MITRE ATT\&CK definitions, producing evidence-grounded, source-anchored TTP extraction outputs.

\section{\tool~Design}
\label{sec:design}

\begin{figure*}[t]
\centering
\includegraphics[width=\textwidth]{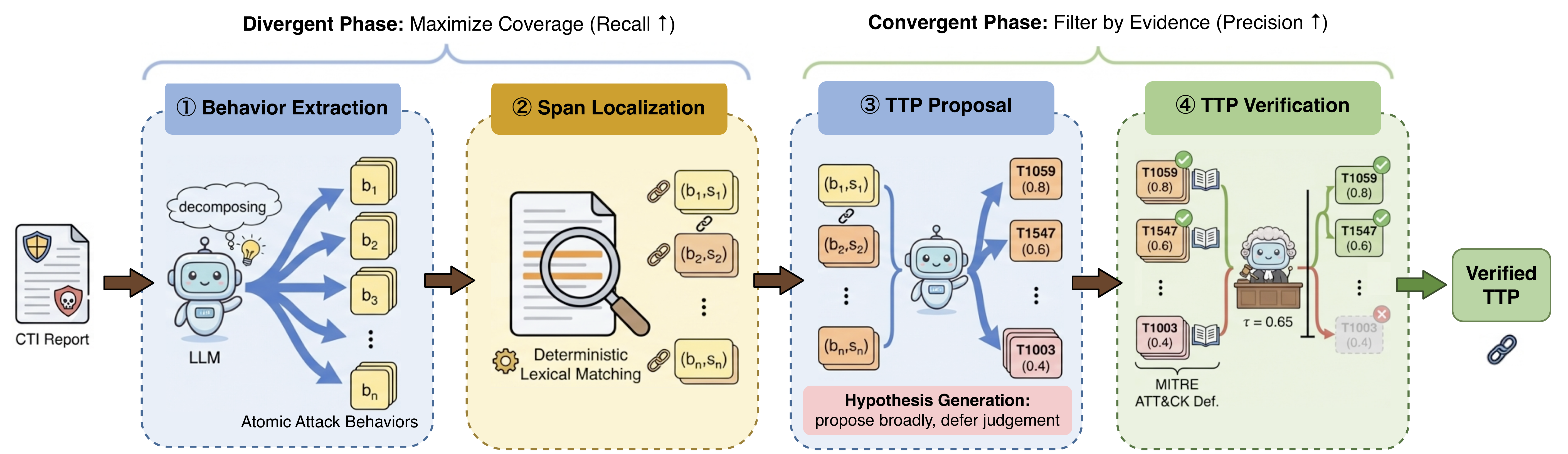}
\caption{\tool~pipeline overview. The pipeline is organized into two phases.
The \textbf{divergent phase} (Stages~\ding{192} and~\ding{194}) maximizes candidate coverage by decomposing a CTI report into atomic attack behaviors and proposing multiple candidate MITRE ATT\&CK techniques per behavior.
The \textbf{convergent phase} (Stages~\ding{193} and~\ding{195}) filters candidates through evidence grounding: span localization anchors each behavior to a specific sentence window via deterministic lexical matching, and TTP verification retains only candidates whose confidence exceeds a threshold $\tau$ after cross-referencing the localized evidence against authoritative MITRE ATT\&CK definitions.}
\label{fig:pipeline}
\end{figure*}

As shown in Figure~\ref{fig:pipeline}, \tool~operationalizes the \emph{diverge-then-converge} principle through a four-stage pipeline organized into two phases. The \textbf{divergent phase}, behavior extraction (\S\ref{sec:behavior}) and TTP proposal (\S\ref{sec:proposal}), prioritizes recall by decomposing the report into atomic behaviors and generating a broad set of candidate techniques. The \textbf{convergent phase}, span localization (\S\ref{sec:span}) and TTP verification (\S\ref{sec:verify}), prioritizes precision by grounding each candidate to localized textual evidence and discarding those that lack support.

\subsection{Behavior Extraction}
\label{sec:behavior}

The first stage decomposes a raw CTI report into a set of atomic attack behaviors. Given an input document $\mathcal{D}$, we prompt an LLM to produce a list of minimal, self-contained descriptions of adversarial actions that are explicitly mentioned in the text. Each behavior $b_i$ isolates a single attack action while explicitly preserving critical technical artifacts such as tool names, command strings, and registry paths. This preservation is essential to prevent \emph{context collapse}, where the LLM abstracts away specific artifacts during extraction, producing descriptions too generic for accurate technique identification.
This decomposition addresses a fundamental challenge in document-level TTP extraction: a single CTI report may describe dozens of distinct attack actions spanning multiple tactics, and directly prompting an LLM to enumerate all relevant techniques from the full document leads to omissions, conflation, and loss of specificity. By isolating individual behaviors, this stage creates atomic units that can each be independently grounded and classified in subsequent stages.
\subsection{Span Localization}
\label{sec:span}

This stage is the architectural keystone of \tool: it bridges the divergent and convergent phases by establishing the extractive evidence link that all downstream predictions must pass through.
For each behavior $b_i \in \mathcal{B}$, a deterministic span localization module identifies the most relevant contiguous sentence window $s_i$ in $\mathcal{D}$ that supports $b_i$. The localization algorithm operates in three steps. First, \emph{bag-of-words overlap scoring}: for each candidate sentence window of length up to $\ell_{\max}$ sentences, we compute the fraction of normalized behavior tokens that appear in the window. Second, \emph{frequency-aware weighting}: tokens that appear in more than a top-$p$ fraction of the corpus are down-weighted to prevent common terms from dominating the score. Third, \emph{n-gram matching}: bigram and trigram overlap is computed as a secondary signal to reward contiguous phrase matches that indicate specific technical content (e.g., tool names, file paths). The window with the highest combined score is selected as the evidence span $s_i$.
Rather than receiving the full document $\mathcal{D}$ as context, the TTP proposal and verification stages receive only the localized span $s_i$. This input-level restriction ensures that proposed techniques are grounded in a specific, identifiable passage.

\subsection{TTP Proposal}
\label{sec:proposal}

Given a behavior $b_i$ and its localized evidence span $s_i$, the TTP proposal stage generates a set of candidate MITRE ATT\&CK techniques. We prompt an LLM to propose up to $k$ parent-level technique identifiers that plausibly correspond to the observed behavior, conditioned on both $b_i$ and $s_i$.
This design draws inspiration from Best-of-N sampling~\citep{gao2023scaling, wu2024inference}, which improves generation quality by producing $N$ independent samples and selecting the highest-scoring one. However, independent sampling tends to cluster around the model's mode, yielding near-duplicate candidates. We instead instruct the LLM to propose $k$ \emph{distinct} techniques within a single generation, encouraging diverse interpretations of the same behavior and improving coverage of the candidate space.
The prompt constrains outputs to valid technique identifiers present in the provided MITRE ATT\&CK corpus. By separating proposal from verification, \tool~avoids the premature filtering that characterizes single-step approaches.

\subsection{TTP Verification}
\label{sec:verify}
The final stage evaluates whether each proposed technique is genuinely supported by the localized evidence. For every candidate technique $t_j$ proposed for behavior $b_i$, the verification module provides the LLM with three inputs: the behavior text $b_i$, its evidence span $s_i$, and the official MITRE ATT\&CK description of technique $t_j$. The model is instructed to determine whether the technique is \emph{explicitly supported} based solely on the provided evidence, without relying on external knowledge or assumptions beyond what is stated in $s_i$. By providing the canonical definition, we standardize the verification criterion: the model assesses whether the evidence span describes an action that matches the official technique semantics. Supporting ATT\&CK updates or an alternative taxonomy requires only refreshing the definition corpus. The verifier outputs a presence confidence score $c_j^{\text{ver}} \in [0, 1]$ for each candidate. Techniques satisfying $c_j^{\text{ver}} \geq \tau$ are retained as final predictions. The threshold $\tau$ serves as an explicit, tunable precision--recall knob. Higher values of $\tau$ yield fewer but more confident predictions (favoring precision), while lower values admit more candidates (favoring recall). We predefined $\tau = 0.7$ as the default threshold, corresponding to the lower bound of ``strong support'' in the verifier’s scoring rubric. 

\section{Evaluation}
\label{sec:eval}

We evaluate \tool~along four dimensions: overall effectiveness against baselines (\S\ref{sec:baseline}), the contribution of individual components (\S\ref{sec:ablation}), robustness across LLM backbones (\S\ref{sec:backbone}), and the verification threshold's impact on precision--recall trade-offs (\S\ref{sec:tau}).

\begin{table*}[t]
\centering
\caption{Document-level results on TRAM-Clean and \tool-Bench. Best results in \textbf{bold}. -- indicates not evaluated.}
\label{tab:main_results}
\small
\renewcommand\arraystretch{1.1}
\setlength{\tabcolsep}{4pt}
\begin{tabular}{l|l|ccc|ccc}
\Xhline{1.2pt}
\rowcolor{tablerow}
 & & \multicolumn{3}{c|}{\textbf{TRAM-Clean}} & \multicolumn{3}{c}{\textbf{\tool-Bench}} \\
\rowcolor{tablerow}
\textbf{Method} & \textbf{Type} & \textbf{Prec.} & \textbf{Rec.} & \textbf{F1} & \textbf{Prec.} & \textbf{Rec.} & \textbf{F1} \\
\Xhline{1.2pt}
\rowcolor{gray!8}
TTPDrill        & Rule  & 1.49 & 5.03 & 2.19 & 4.00  & 35.72 & 7.16  \\
One-Shot GPT-4o & LLM   & 9.56 & 10.51 & 9.66 & 62.99  & 52.80 & 44.23  \\
\rowcolor{gray!8}
CoT GPT-4o      & LLM   & 12.54 & 14.59 & 12.99 & 69.90 & 50.70 & 58.02 \\
CoVe GPT-4o     & LLM   & -- & -- & -- & 67.0 & 54.8 & 58.9 \\
\hline
\rowcolor{gray!8}
\tool            & LLM   & \textbf{74.59} & \textbf{82.10} & \textbf{76.48} & \textbf{82.15} & \textbf{95.50} & \textbf{87.39} \\
\Xhline{1.2pt}
\end{tabular}
\end{table*}

\begin{table*}[t]
\centering
\caption{\tool{} vs.\ CoT by technique document frequency (GPT-4o, \tool-Bench). Scores are macro-averaged over techniques within each frequency bucket.}
\label{tab:rare}
\small
\renewcommand\arraystretch{1.1}
\setlength{\tabcolsep}{4pt}
\begin{tabular}{l|c|ccc|ccc|c}
\Xhline{1.2pt}
\rowcolor{tablerow}
\textbf{Bucket} & \textbf{\#Tech} & \multicolumn{3}{c|}{\textbf{\tool}} & \multicolumn{3}{c|}{\textbf{CoT}} & \textbf{$\Delta$F1} \\
\rowcolor{tablerow}
 & & \textbf{Prec.} & \textbf{Rec.} & \textbf{F1} & \textbf{Prec.} & \textbf{Rec.} & \textbf{F1} & \\
\Xhline{1.2pt}
\rowcolor{gray!8}
Rare ($<$5)      & 66 & 70.4 & 83.1 & 72.6 & 32.3 & 35.1 & 30.8 & +41.8 \\
Medium (5--20)   & 37 & 87.1 & 94.4 & 89.6 & 57.3 & 49.8 & 49.0 & +40.6 \\
\rowcolor{gray!8}
Frequent ($>$20) & 22 & 88.2 & 97.4 & 92.2 & 74.4 & 58.4 & 62.5 & +29.7 \\
\Xhline{1.2pt}
\end{tabular}
\end{table*}

\subsection{Evaluation Setup}

\paragraph{Datasets.}
We evaluate on two complementary datasets. TRAM-Clean is our corrected version of the sentence-level TRAM dataset \citep{tram_dataset}, whose original annotations were designed to fine-tune a BERT-based~\citep{devlin2019bert} NER model and contain substantial false positives and false negatives. \tool-Bench is a new document-level benchmark of 150 full CTI reports drawn from 12 vendor and journalism sources published between 2022 and 2025, each describing a concrete attack campaign in sufficient technical depth for ATT\&CK labeling and absent from existing TTP extraction benchmarks. It requires extracting TTPs directly from complete reports without pre-segmentation. It covers 125 unique MITRE ATT\&CK techniques, 2.5 times the 50 covered by TRAM, including 66 rare techniques (fewer than 5 occurrences) that support long-tail evaluation. Annotations for both datasets are produced by expert annotators working independently, with disagreements adjudicated against the official MITRE ATT\&CK technique descriptions. TRAM-Clean removes unsupported labels and adds missing labels through re-annotation blind to system predictions, and \tool-Bench double annotation reaches Cohen's $\kappa = 0.76$. Full source list, annotation procedure, and inter-annotator agreement are in Appendix~\ref{app:annotation}.

\paragraph{Baselines.}
We compare \tool~against baselines from two methodological families.
For \emph{rule-based} approaches, we include TTPDrill~\citep{ttpdrill}, which uses dependency parsing with a manually constructed threat-action ontology and BM25 scoring. For \emph{LLM-based} approaches, we include two direct prompting baselines: \textbf{One-Shot}, which extracts ATT\&CK techniques from each report, and \textbf{CoT}, which adds explicit chain-of-thought reasoning to the same prompt. We also evaluate factored Chain-of-Verification (CoVe)~\citep{dhuliawala-etal-2024-chain} to test whether adding verification alone accounts for our gains. CoVe generates verification questions and independently answers them before revising its initial response. We conduct this comparison in the full-report setting of \tool-Bench, answering verification questions against the entire report under the same conditions and prompt as CoT. The two single-pass baselines are particularly informative because they isolate the effect of reasoning depth without architectural decomposition, directly testing our hypothesis that single-pass inference couples precision and recall. All baselines are evaluated using official implementations or faithful reproductions.
We exclude supervised neural models as these methods require substantial labeled training data, and our evaluation datasets cannot support a fair comparison under their expected training regime.

\paragraph{Metrics.}
TTP extraction is a multi-label prediction problem in which each report may map to multiple techniques and models must identify the complete set. We evaluate at the \emph{document level} by aggregating predictions across the full report and comparing the resulting technique set to the ground truth. For TRAM-Clean, predictions over its pre-segmented sentences are aggregated per report.
Final scores are macro-averages over all documents: $\overline{\text{F1}} = \frac{1}{|D|} \sum_{d \in D} \text{F1}_d$.

\subsection{RQ1: Overall Effectiveness}
\label{sec:baseline}

Table~\ref{tab:main_results} reports document-level results. \tool~achieves the highest F1 on both datasets by a wide margin: \textbf{87.39}\% on \tool-Bench (+29.37 over CoT GPT-4o at 58.02\%) and \textbf{76.48}\% on TRAM-Clean (+63.49 over CoT GPT-4o at 12.99\%). \tool~improves both precision and recall over CoT on both benchmarks, indicating the gains generalize across source distributions and annotation quality. We quantify the residual run-to-run variability of API serving by repeating \tool~and CoT with the GPT-4o backbone three times each on a fixed 50-report subset. On this subset, F1 is 81.9 ± 1.0\% for \tool~and 55.2 ± 0.1\% for CoT, a gap roughly 27 times the larger standard deviation.

The two single-pass baselines exhibit incomplete behavior extraction and surface similarity mismatch, as discussed in Section~\ref{sec:intro}. CoT improves precision over One-Shot but not recall (50.70\% vs.\ 52.80\%), indicating that deeper reasoning alone does not resolve behavior omissions. On \tool-Bench, CoVe achieves 58.9\% F1 versus CoT's 58.02\%, suggesting that verification alone does not account for \tool's gains. TTPDrill trails all LLM baselines on \tool-Bench. Its manually constructed ontology cannot capture the paraphrases and implicit descriptions common in modern CTI reports~\citep{buechel2025sok}.

Table~\ref{tab:rare} compares \tool~and CoT by technique document frequency. The improvement is even larger for rarer techniques, indicating that the advantage does not rest on frequent techniques but extends across the long tail.

\begin{figure*}[t]
\centering
\begin{minipage}[t]{0.52\textwidth}
    \centering
    \includegraphics[width=\linewidth]{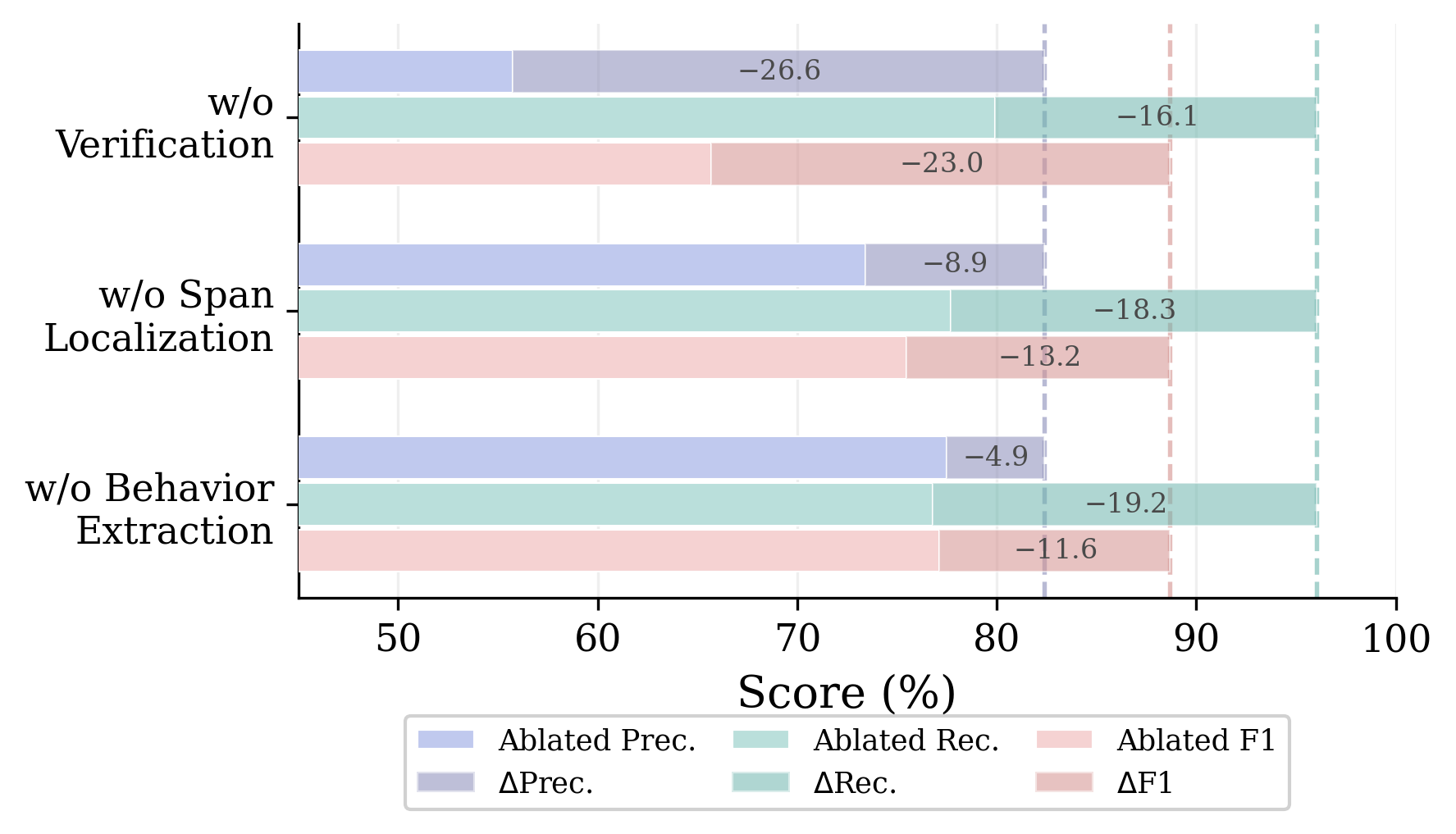}
    \subcaption{Ablation on \tool-Bench (GPT-4o). Darker segments show the drop ($\Delta$) from the full pipeline.}
    \label{fig:ablation}
\end{minipage}
\hfill
\begin{minipage}[t]{0.45\textwidth}
    \centering
    \includegraphics[width=\linewidth]{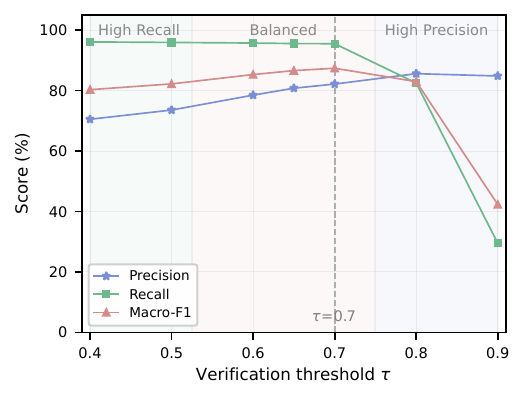}
    \subcaption{Effect of $\tau$ on \tool-Bench (GPT-4o). Macro-F1 peaks at $\tau = 0.7$.}
    \label{fig:tau_sensitivity}
\end{minipage}
\caption{Ablation and threshold sensitivity analysis on \tool-Bench (GPT-4o).}
\label{fig:ablation_and_tau}
\end{figure*}

\subsection{RQ2: Ablation Study}
\label{sec:ablation}

\paragraph{Setup.}
To quantify the contribution of each pipeline stage, we evaluate four variants of \tool~on \tool-Bench: (1) the full pipeline; (2)~\emph{w/o Behavior Extraction}, where the raw document is passed directly to span localization without decomposition; (3)~\emph{w/o Span Localization}, where TTP proposal and verification operate over the full document and atomic behavior rather than a localized evidence window; and (4)~\emph{w/o Verification}, where all proposed candidates above a minimal confidence threshold are accepted without the verification stage.

\paragraph{Results.}
As shown in Figure~\ref{fig:ablation}, each ablation produces a distinct degradation pattern. Removing \emph{behavior extraction} causes the largest recall drop ($\Delta = -19.2\%$, from 95.50\% to 76.30\%) while precision remains relatively stable ($\Delta = -4.9\%$). Without decomposition into atomic units, techniques embedded in complex paragraphs are missed, but the convergent phase still filters unsupported candidates.
Removing \emph{verification} causes the largest precision drop ($\Delta = -26.6\%$, from 82.15\% to 55.55\%) and the largest overall F1 degradation ($\Delta = -23.0\%$), as proposed candidates are accepted without evidence-based scrutiny. Recall also decreases ($\Delta = -16.1\%$), reflecting the loss of MITRE definition cross-referencing that helps disambiguate borderline candidates.
Removing \emph{span localization} degrades both precision and recall substantially ($\Delta = -8.9\%$ and $-18.3\%$, respectively). Without evidence grounding, proposal and verification operate over the full document and atomic behavior, increasing both hallucination risk and context dilution.

\looseness=-1Taken together, the ablation reveals that the stages are not merely additive but \emph{mutually reinforcing}. Verification contributes to recall as well as precision, and span localization benefits not only precision (by excluding irrelevant passages) but also recall (by reducing noise that causes the verifier to reject valid candidates). Evidence grounding makes both upstream proposals and downstream verification more effective. This cross-phase interaction explains why \tool's overall gains exceed the sum of what any single stage contributes in isolation.

\paragraph{Validating faithfulness and traceability.}
Traceability is meaningful only if the intermediate outputs it exposes are faithful. We assess them on 40 randomly sampled \tool-Bench reports (777 behaviors): 99.87\% of localized spans contain the text supporting their behavior, 99.6\% of behaviors are fully faithful to the source report, and none of the 199 preserved technical artifacts is fabricated. Moreover, 97.1\% of ground-truth report--technique pairs are surfaced by at least one behavior's proposals, bounding divergent-phase misses at 2.9\%.

\subsection{RQ3: LLM Backbone Effectiveness}
\label{sec:backbone}

\paragraph{Setup.}
To test whether \tool's gains come from the pipeline architecture rather than a specific LLM, we instantiate the full pipeline with multiple backbones spanning two axes: model family (closed-source vs.\ open-source) and scale. Closed-source models include \textbf{GPT-4o}, \textbf{GPT-4o-mini}, \textbf{Claude Sonnet 4.6}, and \textbf{Gemini 2.5 Flash}. Open-source models include \textbf{Qwen 3 235B} and \textbf{DeepSeek v3.2}. For each backbone, we additionally run the CoT baseline under identical conditions. All pipeline hyperparameters ($\ell_{\max}$, $\tau$, temperature) are held constant across backbones. See Appendix~\ref{app:backbone} for the complete configuration.

\begin{table*}[t]
\centering
\caption{Effect of LLM backbone on \tool-Bench. CoT F1 is the single-pass CoT baseline with the same backbone. $\Delta$F1 is \tool{} minus CoT. Pipeline hyperparameters are held constant.}
\label{tab:backbone}
\small
\renewcommand\arraystretch{1.1}
\setlength{\tabcolsep}{4pt}
\begin{tabular}{l|l|ccccc}
\Xhline{1.2pt}
\rowcolor{tablerow}
\textbf{Backbone} & \textbf{Type} & \textbf{Prec.} & \textbf{Rec.} & \textbf{F1} & \textbf{CoT F1} & \textbf{$\Delta$F1} \\
\Xhline{1.2pt}
\rowcolor{gray!8}
GPT-4o              & Closed, large & 82.15 & 95.50 & 87.39 & 58.02 & $+29.37$ \\
GPT-4o-mini         & Closed, small & 61.31 & 57.49 & 58.01 & 45.00 & $+13.01$ \\
\rowcolor{gray!8}
Claude Sonnet 4.6   & Closed, large & 65.09  & 80.12  & 71.07  & 64.80 & $+6.27$ \\
Gemini 2.5 Flash    & Closed, small & 46.62 & 94.41 & 61.62 & 68.00 & $-6.38$ \\
\hline
\rowcolor{gray!8}
Qwen 3 235B         & Open, large   & 57.81 & 80.91 & 66.59 & 57.88 & $+8.71$ \\
DeepSeek v3.2       & Open, large   & 67.05 & 77.94 & 71.04 & 62.57 & $+8.47$ \\
\Xhline{1.2pt}
\end{tabular}
\end{table*}

\paragraph{Results.}
\looseness=-1Table~\ref{tab:backbone} reports results for each backbone on \tool-Bench. Under the shared default threshold $\tau = 0.7$, \tool~improves over its matched single-pass baseline on five of the six backbones, with same-backbone improvements of at least 6.27 F1 points and up to 29.37 on GPT-4o. GPT-4o-mini with \tool~(58.01\% F1) matches the far larger GPT-4o under CoT (58.02\% F1), indicating that a smaller model \emph{with} the pipeline architecture can substitute for a larger model \emph{without} it.
The exception is Gemini 2.5 Flash, which exhibits high recall (94.41\%) but low precision (46.62\%). Inspection of its persisted intermediate outputs shows that 38.5\% of candidates pass verification, compared with 15.7\% for GPT-4o-mini and 25.3\% for Claude Sonnet 4.6. A post-hoc sensitivity check further shows that raising Gemini’s threshold to $\tau = 0.85$ improves F1 from 61.62\% to 70.9\%. Together, these observations suggest that threshold calibration contributes to its underperformance. The performance of the open-source backbones has an additional practical implication. Organizations with data sovereignty or confidentiality requirements can deploy \tool~with a locally hosted backbone, where improvement persists.

\subsection{RQ4: Threshold Sensitivity}
\label{sec:tau}

Figure~\ref{fig:tau_sensitivity} plots precision, recall, and F1 as a function of $\tau$. Verification scores are produced once per candidate, and sweeping $\tau$ is post-hoc thresholding of these cached scores, with no additional cost.
Three findings emerge. First, the frontier is \emph{smooth and monotonic}. As $\tau$ increases from 0.30 to 0.95, precision rises steadily over most of the range while recall falls, with no abrupt jumps or inversions. Small changes in $\tau$ produce proportional changes in system behavior, making the threshold a reliable and predictable control parameter in practice.
Second, the curve reveals a \emph{clear F1-optimal region}. F1 peaks at the predefined threshold $\tau = 0.7$ (87.39\%). The F1 surface is relatively flat in the neighborhood of the optimum. For $\tau \in [0.55, 0.75]$, F1 remains within approximately 3\% of the peak. This robustness to moderate mis-specification of $\tau$ is a desirable property for practical deployment where per-dataset tuning may not be feasible.
Third, the frontier reveals three natural \emph{operating regimes} for different deployment scenarios (shaded regions in Figure~\ref{fig:tau_sensitivity}). In the \textbf{high-recall regime} ($\tau \leq 0.525$), the system achieves recall above 98\% with precision around 63--72\%, suitable for SOC alert triage where missing a technique is costlier than investigating a false alarm. In the \textbf{balanced regime} ($0.525 < \tau \leq 0.75$), F1 is maximized, with the optimal point at $\tau = 0.7$ delivering 82.15\% precision and 95.50\% recall, appropriate for general-purpose CTI analysis. In the \textbf{high-precision regime} ($\tau > 0.75$), precision exceeds 90\% with recall still above 75\%, appropriate for automated response workflows where every prediction must be highly reliable before triggering defensive actions. Practitioners can select their operating point based on organizational risk tolerance by adjusting a single parameter.

\section{Conclusion and Future Work}
We introduced a diverge-then-converge principle and instantiated it in \tool~for TTP extraction, a four-stage pipeline that decouples recall-oriented candidate generation from precision-oriented evidence-grounded verification.
On TRAM-Clean and \tool-Bench, \tool~achieves 76.48\% and 87.39\% macro-F1 respectively, outperforming chain-of-thought prompting with the same backbone by over 29\% on \tool-Bench while improving both precision and recall simultaneously. Ablations confirm the architecture's design: divergent components control recall, convergent components control precision, and a single threshold~$\tau$ provides continuous, inference-free control over the trade-off. 
Results hold across multiple LLM backbones, including open-source models suitable for air-gapped deployment, suggesting that the architectural benefits extend beyond a single model.
Every prediction is traceable to a localized evidence span, supporting analyst auditability.
We will release both evaluation resources, TRAM-Clean and \tool-Bench, to facilitate research on TTP extraction.

We plan to package \tool~as a modular LLM skill, a self-contained capability that can be invoked by broader LLM-based CTI analysis pipelines as a callable component. This would allow downstream systems performing tasks such as threat actor profiling, attack graph construction, or automated report generation to obtain structured, evidence-grounded TTP extractions, lowering the barrier to integrating high-quality TTP extraction into end-to-end threat intelligence workflows.

\section*{Limitations}
\label{sec:limitations}
\tool~targets English CTI reports published by major security vendors and threat intelligence outlets, which is the dominant medium for technical adversary descriptions and the natural use case for analyst tooling. Extension to non-English CTI or to adjacent document genres such as incident response reports and malware analysis writeups would require source-specific annotation and is left to future work. The verification stage relies on the official MITRE ATT\&CK technique descriptions as its canonical reference. Where the underlying taxonomy itself contains semantically adjacent techniques (e.g., within \emph{Discovery} or \emph{Defense Evasion}), \tool~surfaces candidates faithfully but cannot resolve ambiguity that is structurally present in the source ontology. \tool~incurs higher inference cost and latency than CoT, though the overhead remains manageable for offline, low-throughput CTI analysis (Appendix~\ref{app:cost}). Finally, our evaluation measures extraction quality against expert-annotated ground truth. The impact of \tool~on downstream operational metrics in a live SOC, such as triage time, detection coverage, and attribution confidence, requires controlled deployment studies, which fall outside the scope of a benchmark-driven contribution.

\section*{Ethics Statement}
\tool~assists defensive analysts in identifying adversary behaviors described in publicly published CTI reports authored by security vendors and journalists. The MITRE ATT\&CK framework and the source reports are already accessible to both defenders and adversaries, so the net effect of automation is to reduce the analytical burden on defenders rather than to enable new offensive capabilities. The evaluation datasets contain no private user data, and victim-identifying information from source reports is retained only where required for technical TTP labeling. For organizations handling sensitive threat intelligence, our multi-backbone analysis demonstrates that \tool~retains its performance gains with open-source LLMs hosted locally, removing any need to transmit report content to third-party providers.

\bibliography{references}

@inproceedings{buechel2025sok,
author = {B\"{u}chel, Marvin and Paladini, Tommaso and Longari, Stefano and Carminati, Michele and Zanero, Stefano and Binyamini, Hodaya and Engelberg, Gal and Klein, Dan and Guizzardi, Giancarlo and Caselli, Marco and Continella, Andrea and van Steen, Maarten and Peter, Andreas and van Ede, Thijs},
title = {{SoK}: automated {TTP} extraction from {CTI} reports - are we there yet?},
year = {2025},
isbn = {978-1-939133-52-6},
publisher = {USENIX Association},
address = {USA},
booktitle = {Proceedings of the 34th USENIX Conference on Security Symposium},
articleno = {238},
numpages = {21},
location = {Seattle, WA, USA},
series = {SEC '25}
}

@ARTICLE{mitretrieval,
  author={Huang, Yi-Ting and Vaitheeshwari, R. and Chen, Meng-Chang and Lin, Ying-Dar and Hwang, Ren-Hung and Lin, Po-Ching and Lai, Yuan-Cheng and Wu, Eric Hsiao-Kuang and Chen, Chung-Hsuan and Liao, Zi-Jie and Chen, Chung-Kuan},
  journal={IEEE Transactions on Network and Service Management}, 
  title={{MITREtrieval}: Retrieving {MITRE} Techniques From Unstructured Threat Reports by Fusion of Deep Learning and Ontology}, 
  year={2024},
  volume={21},
  number={4},
  pages={4871-4887},
  doi={10.1109/TNSM.2024.3401200}}

@inproceedings{ttpdrill,
author = {Husari, Ghaith and Al-Shaer, Ehab and Ahmed, Mohiuddin and Chu, Bill and Niu, Xi},
title = {{TTPDrill}: Automatic and Accurate Extraction of Threat Actions from Unstructured Text of {CTI} Sources},
year = {2017},
isbn = {9781450353458},
publisher = {Association for Computing Machinery},
address = {New York, NY, USA},
url = {https://doi.org/10.1145/3134600.3134646},
doi = {10.1145/3134600.3134646},
booktitle = {Proceedings of the 33rd Annual Computer Security Applications Conference},
pages = {103–115},
numpages = {13},
location = {Orlando, FL, USA},
series = {ACSAC '17}
}

@INPROCEEDINGS{extractor,
  author={Satvat, Kiavash and Gjomemo, Rigel and Venkatakrishnan, V.N.},
  booktitle={2021 IEEE European Symposium on Security and Privacy (EuroS\&P)}, 
  title={Extractor: Extracting Attack Behavior from Threat Reports}, 
  year={2021},
  volume={},
  number={},
  pages={598-615},
  doi={10.1109/EuroSP51992.2021.00046}}

@inproceedings{ladder,
author = {Alam, Md Tanvirul and Bhusal, Dipkamal and Park, Youngja and Rastogi, Nidhi},
title = {Looking Beyond {IoCs}: Automatically Extracting Attack Patterns from External {CTI}},
year = {2023},
isbn = {9798400707650},
publisher = {Association for Computing Machinery},
address = {New York, NY, USA},
url = {https://doi.org/10.1145/3607199.3607208},
doi = {10.1145/3607199.3607208},
booktitle = {Proceedings of the 26th International Symposium on Research in Attacks, Intrusions and Defenses},
pages = {92–108},
numpages = {17},
location = {Hong Kong, China},
series = {RAID '23}
}

@inproceedings{cheng2025ctinexusautomaticcyberthreat,
      title={{CTINexus}: Automatic Cyber Threat Intelligence Knowledge Graph Construction Using Large Language Models},
      author={Yutong Cheng and Osama Bajaber and Saimon Amanuel Tsegai and Dawn Song and Peng Gao},
      booktitle={2025 IEEE European Symposium on Security and Privacy (EuroS\&P)},
      year={2025},
      organization={IEEE}
}

@article{attackGplus,
author = {Zhang, Yongheng and Du, Tingwen and Ma, Yunshan and Wang, Xiang and Xie, Yi and Yang, Guozheng and Lu, Yuliang and Chang, Ee-Chien},
title = {{AttacKG+}: Boosting attack graph construction with Large Language Models},
year = {2025},
issue_date = {Mar 2025},
publisher = {Elsevier Advanced Technology Publications},
address = {GBR},
volume = {150},
number = {C},
issn = {0167-4048},
url = {https://doi.org/10.1016/j.cose.2024.104220},
doi = {10.1016/j.cose.2024.104220},
journal = {Comput. Secur.},
month = mar,
numpages = {16}
}

@inproceedings{attackkg,
author = {Li, Zhenyuan and Zeng, Jun and Chen, Yan and Liang, Zhenkai},
title = {{AttacKG}: Constructing Technique Knowledge Graph from Cyber Threat Intelligence Reports},
year = {2022},
isbn = {978-3-031-17139-0},
publisher = {Springer-Verlag},
address = {Berlin, Heidelberg},
url = {https://doi.org/10.1007/978-3-031-17140-6_29},
doi = {10.1007/978-3-031-17140-6_29},
booktitle = {Computer Security – ESORICS 2022: 27th European Symposium on Research in Computer Security, Copenhagen, Denmark, September 26–30, 2022, Proceedings, Part I},
pages = {589–609},
numpages = {21},
location = {Copenhagen, Denmark}
}

@techreport{strom2020mitre,
  title = {{MITRE ATT\&CK}: Design and Philosophy},
  author = {Strom, Blake E. and Applebaum, Andy and Miller, Doug P. and Nickels, Kathryn C. and Pennington, Adam G. and Thomas, Cody B.},
  institution = {The MITRE Corporation},
  year = {2020},
  number = {MP180360R1},
  url = {https://attack.mitre.org/}
}

@misc{tram_dataset,
  author       = {James Ross and Jackie Lasky},
  title        = {Our {TRAM} Large Language Model Automates {TTP} Identification in {CTI} Reports},
  year         = {2023},
  howpublished = {\url{https://medium.com/mitre-engenuity/our-tram-large-language-model-automates-ttp-identification-in-cti-reports-5bc0a30d4567}},
  note         = {Medium article; Center for Threat-Informed Defense / MITRE Engenuity} 
}

@inproceedings{baksi-etal-2025-medcoder,
    title = "{M}ed{C}od{ER}: A Generative {AI} Assistant for Medical Coding",
    author = "Baksi, Krishanu Das  and
      Soba, Elijah  and
      Higgins, John J  and
      Saini, Ravi  and
      Wood, Jaden  and
      Cook, Jane  and
      Scott, Jack I  and
      Pudota, Nirmala  and
      Weninger, Tim  and
      Bowen, Edward  and
      Bhattacharya, Sanmitra",
    editor = "Chen, Weizhu  and
      Yang, Yi  and
      Kachuee, Mohammad  and
      Fu, Xue-Yong",
    booktitle = "Proceedings of the 2025 Conference of the Nations of the Americas Chapter of the Association for Computational Linguistics: Human Language Technologies (Volume 3: Industry Track)",
    month = apr,
    year = "2025",
    address = "Albuquerque, New Mexico",
    publisher = "Association for Computational Linguistics",
    url = "https://aclanthology.org/2025.naacl-industry.37/",
    doi = "10.18653/v1/2025.naacl-industry.37",
    pages = "449--459",
    ISBN = "979-8-89176-194-0"
}

@inproceedings{dhuliawala-etal-2024-chain,
    title = "Chain-of-Verification Reduces Hallucination in Large Language Models",
    author = "Dhuliawala, Shehzaad  and
      Komeili, Mojtaba  and
      Xu, Jing  and
      Raileanu, Roberta  and
      Li, Xian  and
      Celikyilmaz, Asli  and
      Weston, Jason",
    editor = "Ku, Lun-Wei  and
      Martins, Andre  and
      Srikumar, Vivek",
    booktitle = "Findings of the Association for Computational Linguistics: ACL 2024",
    month = aug,
    year = "2024",
    address = "Bangkok, Thailand",
    publisher = "Association for Computational Linguistics",
    url = "https://aclanthology.org/2024.findings-acl.212/",
    doi = "10.18653/v1/2024.findings-acl.212",
    pages = "3563--3578"
}

@inproceedings{clh_icd,
    title = "Code Like Humans: A Multi-Agent Solution for Medical Coding",
    author = "Motzfeldt, Andreas Geert  and
      Edin, Joakim  and
      Christensen, Casper L.  and
      Hardmeier, Christian  and
      Maal{\o}e, Lars  and
      Rogers, Anna",
    editor = "Christodoulopoulos, Christos  and
      Chakraborty, Tanmoy  and
      Rose, Carolyn  and
      Peng, Violet",
    booktitle = "Findings of the Association for Computational Linguistics: EMNLP 2025",
    month = nov,
    year = "2025",
    address = "Suzhou, China",
    publisher = "Association for Computational Linguistics",
    url = "https://aclanthology.org/2025.findings-emnlp.1231/",
    doi = "10.18653/v1/2025.findings-emnlp.1231",
    pages = "22612--22627",
    ISBN = "979-8-89176-335-7"
}

@inproceedings{madaan2023selfrefine,
author = {Madaan, Aman and Tandon, Niket and Gupta, Prakhar and Hallinan, Skyler and Gao, Luyu and Wiegreffe, Sarah and Alon, Uri and Dziri, Nouha and Prabhumoye, Shrimai and Yang, Yiming and Gupta, Shashank and Majumder, Bodhisattwa Prasad and Hermann, Katherine and Welleck, Sean and Yazdanbakhsh, Amir and Clark, Peter},
title = {SELF-REFINE: iterative refinement with self-feedback},
year = {2023},
publisher = {Curran Associates Inc.},
address = {Red Hook, NY, USA},
booktitle = {Proceedings of the 37th International Conference on Neural Information Processing Systems},
articleno = {2019},
numpages = {61},
location = {New Orleans, LA, USA},
series = {NIPS '23}
}

@inproceedings{huang2024selfcorrection,
  title={Large Language Models Cannot Self-Correct Reasoning Yet},
  author = {Huang, Jie and Chen, Xinyun and Mishra, Swaroop and Zheng, Huaixiu Steven and Yu, Adams Wei and Song, Xinying and Zhou, Denny},
  booktitle={The Twelfth International Conference on Learning Representations (ICLR)},
  year={2024}
}

@inproceedings{swarup-etal-2025-llm4re,
    title = "{LLM}4{RE}: A Data-centric Feasibility Study for Relation Extraction",
    author = "Swarup, Anushka  and
      Pan, Tianyu  and
      Wilson, Ronald  and
      Bhandarkar, Avanti  and
      Woodard, Damon",
    editor = "Rambow, Owen  and
      Wanner, Leo  and
      Apidianaki, Marianna  and
      Al-Khalifa, Hend  and
      Eugenio, Barbara Di  and
      Schockaert, Steven",
    booktitle = "Proceedings of the 31st International Conference on Computational Linguistics",
    month = jan,
    year = "2025",
    address = "Abu Dhabi, UAE",
    publisher = "Association for Computational Linguistics",
    url = "https://aclanthology.org/2025.coling-main.447/",
    pages = "6670--6691"
}

@inproceedings{bhattacharyya-etal-2025-information,
    title = "Information Extraction from Visually Rich Documents using {LLM}-based Organization of Documents into Independent Textual Segments",
    author = "Bhattacharyya, Aniket  and
      Tripathi, Anurag  and
      Das, Ujjal  and
      Karmakar, Archan  and
      Pathak, Amit  and
      Gupta, Maneesh",
    editor = "Che, Wanxiang  and
      Nabende, Joyce  and
      Shutova, Ekaterina  and
      Pilehvar, Mohammad Taher",
    booktitle = "Proceedings of the 63rd Annual Meeting of the Association for Computational Linguistics (Volume 1: Long Papers)",
    month = jul,
    year = "2025",
    address = "Vienna, Austria",
    publisher = "Association for Computational Linguistics",
    url = "https://aclanthology.org/2025.acl-long.844/",
    doi = "10.18653/v1/2025.acl-long.844",
    pages = "17241--17256",
    ISBN = "979-8-89176-251-0"
}

@article{aggarwal2025fiscalie,
  title={Information Extraction From Fiscal Documents Using {LLMs}},
  author={Aggarwal, Vikram and Kulkarni, Jay and Mascarenhas, Aditi and Narang, Aakriti and Raman, Siddarth and Shah, Ajay and Thomas, Susan},
  journal={arXiv preprint arXiv:2511.10659},
  year={2025}
}

@inproceedings{gao2023scaling,
  title={Scaling laws for reward model overoptimization},
  author={Gao, Leo and Schulman, John and Hilton, Jacob},
  booktitle={International Conference on Machine Learning},
  pages={10835--10866},
  year={2023},
  organization={PMLR}
}

@article{wu2024inference,
  title={Inference scaling laws: An empirical analysis of compute-optimal inference for problem-solving with language models},
  author={Wu, Yangzhen and Sun, Zhiqing and Li, Shanda and Welleck, Sean and Yang, Yiming},
  journal={arXiv preprint arXiv:2408.00724},
  year={2024}
}

@misc{devlin2019bert,
      title={{BERT}: Pre-training of Deep Bidirectional Transformers for Language Understanding}, 
      author={Jacob Devlin and Ming-Wei Chang and Kenton Lee and Kristina Toutanova},
      year={2019},
      eprint={1810.04805},
      archivePrefix={arXiv},
      primaryClass={cs.CL},
      url={https://arxiv.org/abs/1810.04805}, 
}

\appendix
\newpage

\section{Prompt Templates}
\label{app:prompts}

This appendix documents the complete prompt templates used in \tool~and its baselines. All templates use Jinja2 syntax; variables enclosed in \texttt{\{\{ \}\}} are substituted at runtime. We organize prompts by pipeline stage (\S\ref{app:prompt_behavior}--\S\ref{app:prompt_verify}) and include baseline prompts (\S\ref{app:prompt_baseline}--\S\ref{app:prompt_baseline_cot}) for reproducibility. Throughout this appendix, \colorbox{pastelbluebg}{\strut\small system prompts} are shown in blue, \colorbox{pastelgreenbg}{\strut\small user prompts} in green.

\subsection{Baseline: One-Shot GPT-4o}
\label{app:prompt_baseline}

\begin{systempromptbox}[System Prompt --- One-Shot GPT-4o]
\begin{lstlisting}[style=promptstyle]
You are a cybersecurity analyst extracting MITRE ATT&CK techniques from a CTI report.

Rules:
- Extract every ATT&CK technique that is explicitly stated or strongly supported by the report content.
- You must strictly base your judgment on the report.
- Output only ATT&CK technique IDs in parent-technique format: TXXXX, no child technique like TXXXX.XXX.
- You should select TTP IDs that are valid MITRE ATT&CK technique IDs
- Deduplicate technique IDs.
- If no supported techniques are present, return an empty list.

Return strict JSON with this schema:
{
  "ttp_ids": ["TXXXX", "TYYYY"]
}
\end{lstlisting}
\end{systempromptbox}

\begin{userpromptbox}[User Prompt --- One-Shot GPT-4o]
\begin{lstlisting}[style=promptstyle]
Extract all MITRE ATT&CK technique IDs from the following CTI report.

Return only a JSON object that matches the required schema.

CTI report:
{{ report_text }}
\end{lstlisting}
\end{userpromptbox}

\subsection{Baseline: CoT GPT-4o}
\label{app:prompt_baseline_cot}

\begin{systempromptbox}[System Prompt --- CoT GPT-4o]
\begin{lstlisting}[style=promptstyle]
You are a cybersecurity analyst extracting MITRE ATT&CK techniques from a CTI report.

Use a careful step-by-step reasoning process:
1. Identify concrete attacker behaviors described in the report.
2. Map each supported behavior to the most appropriate ATT&CK technique.
3. Check whether there are any omitted TTPs that can be strongly supported by the report.
4. Remove unsupported mappings and deduplicate technique IDs.

Rules:
- Extract every ATT&CK technique that is explicitly stated or strongly supported by the report content.
- You must strictly base your judgment on the report.
- Output only ATT&CK technique IDs in parent-technique format: TXXXX, no child technique like TXXXX.XXX.
- You should select TTP IDs that are valid MITRE ATT&CK technique IDs
- Deduplicate technique IDs.
- If no supported techniques are present, return an empty list.

Return strict JSON with this schema:
{
  "reasoning_steps": [
    {
      "behavior": "short behavior summary",
      "evidence": "brief supporting evidence from the report",
      "mapped_ttp": "TXXXX"
    }
  ],
  "ttp_ids": ["TXXXX", "TYYYY"]
}
\end{lstlisting}
\end{systempromptbox}

\begin{userpromptbox}[User Prompt --- CoT GPT-4o]
\begin{lstlisting}[style=promptstyle]
Read the CTI report below and reason step by step before producing the final ATT&CK technique list.

Requirements:
- Only include techniques supported by the report.
- Normalize all outputs to parent-technique IDs in the form TXXXX.
- Return only a JSON object matching the required schema.

CTI report:
{{ report_text }}
\end{lstlisting}
\end{userpromptbox}

\subsection{Stage~\ding{192}: Behavior Extraction}
\label{app:prompt_behavior}

\begin{systempromptbox}[System Prompt --- Behavior Extraction]
\begin{lstlisting}[style=promptstyle]
You are a cybersecurity expert. Your task is to accurately extract attack behaviors from CTI reports.

Requirements:
1. Each extracted attack behavior must be a very brief sentence, closely matching the original wording
2. Reuse technical terms and verb phrases from the original text as much as possible
3. If the original text contains specific strings (domains, hashes, paths, command lines), prioritize preserving them in behavior_text
4. Each behavior corresponds to one or more sentences in the original text
5. You must output strict JSON format, no explanatory text allowed, no extra fields allowed

Output format (strict JSON, no other text):
{
  "behaviors": [
    {
      "behavior_text": "A brief sentence describing the attack behavior"
    }
  ]
}
\end{lstlisting}
\end{systempromptbox}

\begin{userpromptbox}[User Prompt --- Behavior Extraction]
\begin{lstlisting}[style=promptstyle]
Please extract all attack behaviors from the following CTI report:

{{ raw_text }}

Please output strict JSON format according to the requirements, do not add any explanatory text.
\end{lstlisting}
\end{userpromptbox}

\subsection{Stage~\ding{194}: TTP Proposal}
\label{app:prompt_proposal}

\begin{systempromptbox}[System Prompt --- TTP Proposal]
\begin{lstlisting}[style=promptstyle]
You are a cybersecurity expert. Your task is to recommend the most relevant TTPs (Techniques only, NOT sub-techniques) from the MITRE ATT&CK framework based on the given attack behavior and evidence sentences.

Requirements:
1. You must output strict JSON format, no explanatory text allowed
2. You must output exactly N different TTP candidates (N={{ n }})
3. Each candidate must contain ttp_id (format like "T1059", NOT "T1059.003") and propose_confidence (a float between 0-1)
4. IMPORTANT: You must ONLY use parent techniques (e.g., "T1059"), NOT sub-techniques (e.g., "T1059.003"). Sub-techniques are NOT allowed.
5. You should select TTP IDs that are valid MITRE ATT&CK technique IDs (parent techniques only)

Output format (strict JSON, no other text):
{
  "ttp_candidates": [
    {
      "ttp_id": "T1059",
      "propose_confidence": 0.85
    },
    {
      "ttp_id": "T1071",
      "propose_confidence": 0.75
    }
  ]
}
\end{lstlisting}
\end{systempromptbox}

\begin{userpromptbox}[User Prompt --- TTP Proposal]
\begin{lstlisting}[style=promptstyle]
Recommend {{ n }} MITRE ATT&CK TTP IDs.

Behavior: {{ behavior_text }}

Evidence: {{ evidence_sentences }}

Output JSON (only parent techniques, NOT sub-techniques):
{
  "ttp_candidates": [
    {"ttp_id": "T1059", "propose_confidence": 0.85},
    {"ttp_id": "T1071", "propose_confidence": 0.75}
  ]
}
\end{lstlisting}
\end{userpromptbox}

\subsection{Stage~\ding{195}: TTP Verification}
\label{app:prompt_verify}

\begin{systempromptbox}[System Prompt --- TTP Verification]
\begin{lstlisting}[style=promptstyle]
You are a cybersecurity expert. Your task is to determine whether each MITRE ATT&CK TTP is present/supported in the evidence based on the given attack behavior and evidence sentences.

Important constraints:
1. You must strictly base your judgment on the evidence sentences (best_span_sentences) and TTP definition, and you are not allowed to make inferences based on common knowledge
2. If the key points described in the TTP's definition are not clearly reflected in the evidence sentences, the confidence must be low (close to 0)
3. You must output strict JSON format, no explanatory text allowed
4. Each TTP's presence_confidence must be a float between [0,1]
5. IMPORTANT: Some of the given TTPs may be misleading or not actually present in the evidence. You must carefully and accurately judge whether each TTP is truly supported by the evidence sentences, avoiding false positives.

Confidence calibration rules (required):
- 0.00-0.09: No supporting evidence, or evidence contradicts the TTP definition.
- 0.10-0.29: Very weak support (only vague/indirect lexical overlap, missing core TTP behavior).
- 0.30-0.49: Partial support (some relevant action is present, but key required elements are missing).
- 0.50-0.69: Moderate support (core behavior is plausible in evidence, but important ambiguity remains).
- 0.70-0.89: Strong support (The behavior described in the evidence sentence is consistent with the TTP definition, but the evidence may lack a small number of characteristics present in the definition).
- 0.90-1.00: Very strong support (The behavior described in the evidence sentence is exactly the same as the TTP definition, with little differences).

Scoring discipline:
- Base confidence on evidence quality and completeness, not on prior likelihood.
- Prefer conservative scoring when evidence is ambiguous.
- Use two-decimal precision (e.g., 0.73).

Output format (strict JSON, no other text):
{
  "ttp_verifications": [
    {
      "ttp_id": "T1059",
      "presence_confidence": 0.75
    }
  ]
}
\end{lstlisting}
\end{systempromptbox}

\begin{userpromptbox}[User Prompt --- TTP Verification]
\begin{lstlisting}[style=promptstyle]
Based on the following attack behavior and evidence sentences, determine whether each TTP is supported in the evidence:

Attack Behavior: {{ behavior_text }}

Evidence sentences (you must strictly base your judgment on these sentences):
{{ evidence_sentences }}

TTPs to verify:
{% for ttp_info in ttp_info_list %}
{{ loop.index }}. TTP ID: {{ ttp_info.ttp_id }}
   definition:
   {{ ttp_info.contents_preview }}
{% endfor %}

Please output presence_confidence for each TTP, indicating the degree to which the TTP is supported in the evidence sentences.
\end{lstlisting}
\end{userpromptbox}

\section{Dataset Annotation Protocol}
\label{app:annotation}

This appendix provides the full annotation methodology for both evaluation
resources introduced in Section~\ref{sec:eval}: the cleaned TRAM benchmark
(TRAM-Clean) and the new \tool-Bench dataset. Section~\ref{app:tram_cleaning}
describes the systematic cleaning protocol applied to the original TRAM
annotations. Section~\ref{app:bench_protocol} describes the end-to-end
protocol used to construct \tool-Bench from scratch.
Section~\ref{app:iaa} reports inter-annotator agreement statistics for both
datasets.

\paragraph{Licensing.}
The TRAM repository is distributed under the Apache 2.0 license, while MITRE ATT\&CK is available under MITRE’s terms of use, which permit research use and require preservation of copyright and license notices in redistributed copies. The original CTI reports remain subject to their publishers’ copyright and applicable terms of use.

\subsection{TRAM-Clean: Annotation Cleaning Protocol}
\label{app:tram_cleaning}

\paragraph{Background and motivation.}
The TRAM dataset~\citep{tram_dataset} consists of 150 CTI reports with
sentence-level annotations covering the 50 most frequently observed MITRE
ATT\&CK techniques. Its annotations were originally produced to fine-tune a
BERT-based NER model and were not designed as a gold-standard evaluation
benchmark. Our preliminary review identified two systematic classes of
annotation error that materially distort evaluation:

\begin{itemize}
    \item \textbf{False positives} (over-annotation): technique labels
    assigned to sentences that describe generic system activity,
    reconnaissance context, or background narrative rather than a concrete
    attacker action matching the labeled technique's definition. These inflate
    measured precision and artificially penalize conservative models.

    \item \textbf{False negatives} (under-annotation): sentences that clearly
    describe a MITRE ATT\&CK technique but carry no annotation, typically
    because the technique is expressed through paraphrase, implicit context,
    or tool-specific terminology rather than the keyword patterns that
    dominated the original annotation pass. These depress measured precision and
    disadvantage models with strong contextual understanding.
\end{itemize}

\paragraph{Annotator selection and training.}
All cleaning was performed by 3 annotators with verifiable
backgrounds in both cybersecurity operations and the MITRE ATT\&CK framework
(threat intelligence analysts with operational SOC experience and familiarity
with the MITRE ATT\&CK knowledge base). Before the main annotation, annotators
completed a calibration phase on a held-out set of 10 reports not drawn from
TRAM, during which disagreements were discussed and resolved to establish
shared interpretation norms. Annotators were explicitly instructed to use the
\textbf{official MITRE ATT\&CK technique descriptions} as the sole reference
for deciding whether a technique label is warranted; colloquial associations
or keyword matches were not considered sufficient justification. All annotators were informed of and consented to the research use and public release of their annotations.

\paragraph{Blind protocol.}
To prevent any system's predictions from influencing ground-truth corrections,
annotators had no access to the output of any TTP extraction system, including
\tool, during the cleaning process. The only reference materials permitted
were the original CTI report and the ATT\&CK knowledge base.

\paragraph{Procedure.}
The quality audit proceeded in three passes.

\begin{enumerate}
    \item \textbf{Independent review.} Each annotator independently reviewed
    every report in the original TRAM dataset, assessing both report-level
    quality (sufficient technical depth, coherent narrative, absence of
    systematic annotation artifacts) and label-level correctness. For each
    sentence--technique pair, the annotator recorded one of three decisions:
    \textsc{Keep} (the annotation is correct), \textsc{Remove} (false
    positive), or \textsc{Flag} (uncertain, requires discussion). Annotators
    also recorded any unannotated sentences they believed should carry a label
    (\textsc{Add}). Reports with pervasive annotation errors or insufficient
    technical content were flagged for exclusion.

    \item \textbf{Adjudication.} All report-level exclusion decisions, all
    \textsc{Flag} decisions, all \textsc{Remove} decisions where annotators
    disagreed, and all proposed \textsc{Add} cases were brought to a joint
    adjudication session. Each contested case was resolved by majority vote
    among the three annotators, with the requirement that the deciding vote be
    accompanied by a citation to the relevant passage in the official MITRE
    ATT\&CK technique description.

    \item \textbf{Consistency audit.} A final pass checked that all
    annotations for the same technique were applied consistently across the
    retained reports (e.g., if a particular phrasing was accepted in one
    report, equivalent phrasings in other reports were audited for
    corresponding labels).
\end{enumerate}

\paragraph{Scope and outcome.}
The quality audit was applied to a subset of the original TRAM dataset. Due to the labor-intensive nature of manual annotation review, we sampled 21 reports from the full 150-report corpus for cleaning, prioritizing reports that span diverse threat actors, attack campaigns, and technique distributions to ensure broad coverage. After cleaning, these 21 reports yield 2{,}066 sentences with 665 technique labels across 37 unique techniques, constituting the TRAM-Clean evaluation set.
Table~\ref{tab:tram_cleaning_stats} summarizes the composition of TRAM-Clean.

\begin{table}[h]
\centering
\caption{Composition of TRAM-Clean after quality audit of the original TRAM dataset.}
\label{tab:tram_cleaning_stats}
\small
\begin{tabular}{lr}
\toprule
\textbf{Statistic} & \textbf{Value} \\
\midrule
Sentences in TRAM-Clean         & 2{,}066 \\
Technique labels in TRAM-Clean  & 665 \\
Unique techniques               & 37 \\
Avg.\ sentences per report      & 98.4 \\
Avg.\ labels per sentence       & 1.23 \\
\bottomrule
\end{tabular}
\end{table}

\subsection{\tool-Bench: New Dataset Annotation Protocol}
\label{app:bench_protocol}

\paragraph{Source selection.}
\tool-Bench was constructed to complement TRAM-Clean by providing evaluation
data with higher diversity and no legacy annotation noise. We collected
150 CTI reports published between 2022 and
2025 from 12 sources, including
vendor threat intelligence blogs (Palo Alto Networks Unit 42, CrowdStrike, Bitdefender Labs, Bitdefender Business Insights, ESET WeLiveSecurity, Zscaler ThreatLabz, Forcepoint X-Labs, LevelBlue SpiderLabs), infrastructure provider security blogs (Cloudflare), and independent security journalism outlets (Krebs on Security, The Hacker News, Security.com Threat Intelligence). Source selection
criteria were: (1) the report describes a concrete attack campaign or
adversary behavior at sufficient technical depth for MITRE ATT\&CK labeling;
(2) the report has not appeared in any existing TTP extraction benchmark;
and (3) the source covers a range of adversary sophistication levels,
targeted sectors, and geographic contexts to avoid systematic bias.

Reports shorter than 15 sentences or longer than
300 sentences were excluded to control for length-related
variability in evaluation metrics.

\paragraph{Preprocessing.}
Each report was segmented into sentences using a rule-based sentence splitter
calibrated on cybersecurity text (handling abbreviations such as ``e.g.'',
``Fig.'', and version strings that trigger false sentence boundaries in generic
tools). Code blocks, tables, and structured IoC lists were preserved as single
units and treated as individual ``sentences'' for annotation purposes, since
they frequently encode technique-relevant evidence. Document metadata
(title, publication date, author) was stripped before annotation to prevent
anchoring bias.

\paragraph{Annotation unit.}
Annotators labeled at the \textbf{sentence level}. For each sentence, they
assigned zero or more parent-level MITRE ATT\&CK technique IDs (format:
\texttt{TXXXX}). Sub-techniques (\texttt{TXXXX.XXX}) were not used. If a
sub-technique was the most precise match, annotators recorded the
corresponding parent technique. This design is consistent with TRAM and
enables fair comparison across datasets.

\paragraph{Annotator pool and qualification.}
3 annotators participated in \tool-Bench annotation.
Each annotator satisfied the following criteria:
\begin{itemize}
    \item Holds a cybersecurity qualification (GCTI, OSCP, or equivalent industry experience in threat intelligence analysis) or has authored peer-reviewed research
    in threat intelligence or intrusion analysis.
    \item Has demonstrated familiarity with the MITRE ATT\&CK framework
    assessed by a qualification test requiring correct labeling of
    30 curated example sentences at $\geq$80\%
    accuracy.
\end{itemize}
Annotators were compensated at standard institutional rates appropriate for expert annotation work in their region of residence and were not involved
in the development of \tool. All annotators were informed of and consented to the research use and public release of their annotations.

\paragraph{Annotation procedure.}
The annotation process followed a four-phase workflow:

\begin{enumerate}
    \item \textbf{Guideline review.} Before annotation began, all annotators
    received a 20-page annotation guideline document covering: (a) the scope
    and structure of the MITRE ATT\&CK framework; (b) decision criteria for
    assigning vs.\ withholding technique labels; (c) worked examples of
    boundary cases (implicit descriptions, generic system activity, compound
    behaviors); and (d) instructions for handling uncertain cases.

    \item \textbf{Pilot round.} Each annotator independently labeled a
    common set of 5 pilot reports. Pilot annotations were compared to a
    consensus set prepared by the authors, and annotators with agreement below
    0.60 Cohen's $\kappa$ on the pilot received targeted feedback
    before proceeding to the main annotation.

    \item \textbf{Double-annotation.} Each report in the main annotation batch
    was independently labeled by exactly two annotators. Annotators worked
    without visibility into each other's labels.

    \item \textbf{Conflict resolution.} Sentence--technique pairs where the
    two annotators disagreed were resolved by a third annotator acting as
    adjudicator. The adjudicator had access to both annotators' labels but not
    their rationales, and was required to produce a decision with an explicit
    citation to the relevant ATT\&CK technique description. If the adjudicator
    disagreed with both primary annotators, all three labels were reviewed
    jointly and resolved by majority vote.
\end{enumerate}

\paragraph{Annotation interface.}
Annotation was conducted in a custom web interface built on top of
Label Studio. The interface displayed the full
report alongside the ATT\&CK technique browser, allowing annotators to look
up technique definitions without leaving the annotation environment. Each
sentence was presented in its document context (three preceding and three
following sentences were visible but not annotatable) to assist with
coreference resolution and implicit behavior identification.

\paragraph{Quality control.}
To detect annotation drift over time, each annotator re-labeled a random
sample of 5\% of their previously completed sentences midway through the
annotation campaign (cross-over check). Labels that differed from the
original by more than 10\% were flagged for review. In
addition, 15 ``honeypot'' sentences with known ground truth
(verified by three senior annotators) were inserted into each annotator's
queue. Any annotator whose honeypot accuracy fell below
85\% had their batch re-reviewed.

\subsection{Inter-Annotator Agreement}
\label{app:iaa}

We report inter-annotator agreement (IAA) using two complementary metrics.
\textbf{Cohen's $\kappa$} measures pairwise agreement on the binary
presence/absence decision for each technique at the sentence level, accounting
for chance agreement. \textbf{Macro-averaged F1} treats one annotator's
labels as predictions and the other's as ground truth, averaged over all
technique classes with at least one occurrence. This is more informative than
$\kappa$ for highly imbalanced label distributions typical of CTI annotation.

Table~\ref{tab:iaa} reports IAA for both datasets. For TRAM-Clean, we report
the pairwise $\kappa$ between annotators computed over the set of contested
cases (i.e., cases that reached the adjudication stage). For \tool-Bench, we
report the average pairwise $\kappa$ over all double-annotated pairs prior to
adjudication.

\begin{table*}[t]
\centering
\caption{Inter-annotator agreement for TRAM-Clean and \tool-Bench. $\kappa \geq 0.60$ is conventionally interpreted as substantial agreement.}
\label{tab:iaa}
\small
\begin{tabular}{lccc}
\toprule
\textbf{Dataset} & \textbf{Cohen's $\kappa$} & \textbf{Macro-F1} & \textbf{Scope} \\
\midrule
TRAM-Clean (contested cases)    & 0.74 & 78.3 & 3 pairs \\
\tool-Bench (pre-adjudication)  & 0.76 & 81.2 & 3 pairs \\
\bottomrule
\end{tabular}
\end{table*}

Both datasets achieve substantial inter-annotator agreement ($\kappa > 0.70$), confirming that the annotation task is well-defined and reproducible under our protocol. The slightly higher agreement on \tool-Bench ($\kappa = 0.76$) compared to TRAM-Clean contested cases ($\kappa = 0.74$) is expected. TRAM-Clean agreement is measured only over cases that reached adjudication (i.e., the hardest subset), while \tool-Bench reports pre-adjudication agreement across all double-annotated pairs. The macro-F1 agreement (78.3\% and 81.2\%) provides additional assurance that consistency holds across rare techniques, not just frequent ones, an important property given that \tool-Bench covers 66 techniques with fewer than 5 occurrences.

We also analyzed disagreement patterns to identify systematically difficult
annotation cases. The most common sources of disagreement were:

\begin{itemize}
    \item \textbf{Compound behaviors}: sentences describing two distinct
    attack actions (e.g., credential access followed by lateral movement)
    where annotators differed on which technique was primary.

    \item \textbf{Implicit techniques}: sentences that strongly imply a
    technique through contextual inference (e.g., describing the \emph{effect}
    of a persistence mechanism without naming the mechanism itself), where
    annotators differed on whether implication was sufficient for labeling.

    \item \textbf{Tactic--technique boundary}: sentences that clearly indicate
    a tactic (e.g., exfiltration) but do not specify the technique used,
    leading to disagreement between annotators who labeled the most likely
    technique and those who withheld the label for lack of explicit support.
\end{itemize}

These patterns motivated the annotation guideline's explicit treatment of each
case and informed the calibration examples used in annotator training.

\section{LLM Backbone Configuration}
\label{app:backbone}

Table~\ref{tab:backbone_config} summarizes the LLM backbones used in the multi-backbone analysis (\S\ref{sec:backbone}). For each model, we report the API identifier or model version used in our experiments, the architecture type, publicly available parameter counts, and context window size.

\begin{table*}[t]
\centering
\caption{LLM backbone specifications. ``Context'' denotes the maximum context window in tokens. MoE = Mixture-of-Experts. Active parameters are those used per forward pass.}
\label{tab:backbone_config}
\small
\renewcommand\arraystretch{1.1}
\setlength{\tabcolsep}{4pt}
\begin{tabular}{llllr}
\Xhline{1.2pt}
\rowcolor{tablerow}
\textbf{Backbone} & \textbf{API Identifier} & \textbf{Architecture} & \textbf{Parameters} & \textbf{Context} \\
\Xhline{1.2pt}
\rowcolor{gray!8}
GPT-4o            & \texttt{gpt-4o}          & Dense       & Undisclosed         & 128K \\
GPT-4o-mini       & \texttt{gpt-4o-mini}     & Dense       & Undisclosed         & 128K \\
\rowcolor{gray!8}
Claude Sonnet 4.6 & \texttt{claude-sonnet-4-6} & Dense       & Undisclosed         & 200K \\
Gemini 2.5 Flash  & \texttt{gemini-2.5-flash}           & Dense       & Undisclosed         & 1M \\
\hline
\rowcolor{gray!8}
Qwen 3 235B       & \texttt{qwen3-235b-a22b}            & MoE         & 235B (22B active)   & 131K \\
DeepSeek v3.2     & \texttt{deepseek-chat (v3)}         & MoE         & 671B (37B active)   & 128K \\
\Xhline{1.2pt}
\end{tabular}
\end{table*}

\paragraph{Shared hyperparameters.}
All pipeline hyperparameters are held constant across backbones to isolate the effect of the underlying model. We set the span localization window $\ell_{\max} = 5$ sentences, the
number of TTP proposals per behavior $k = 5$, and the corpus frequency cutoff for token down-weighting at the top-$p = 0.01$ fraction. The verification threshold is $\tau = 0.7$ for all main comparisons across datasets and backbones. For LLM inference, we set temperature to 0 where supported and retain its default otherwise. All other API parameters (e.g., top-$p$) remain at provider defaults.

\begin{table}[ht]
\centering
\caption{Shared pipeline hyperparameters used across all backbone experiments.}
\label{tab:hyperparams}
\small
\begin{tabular}{lr}
\toprule
\textbf{Hyperparameter} & \textbf{Value} \\
\midrule
Temperature                          & 0.0 \\
Verification threshold $\tau$        & 0.7 \\
Max span window $\ell_{\max}$ (sentences) & 5 \\
TTP proposals per behavior $k$       & 5 \\
Top-$p$ frequency cutoff             & 0.01 \\
\bottomrule
\end{tabular}
\end{table}

\noindent All closed-source models were accessed via their respective commercial APIs. Open-source models (Qwen 3 235B and DeepSeek v3.2) were accessed via third-party API providers. As noted in Section~\ref{sec:backbone}, these models can alternatively be deployed locally for organizations with data sovereignty requirements.

\section{Case Study of Components}
\label{app:qualitative}

The following case studies illustrate the effects of individual components.

\paragraph{Behavior extraction.}
In the DragonRank report, a paragraph describes cloning an administrator's permissions to a guest account and then executing a credential-dumping tool. The \textit{w/o Behavior Extraction} ablation identifies credential dumping but misses the account-manipulation technique associated with the permission change. By isolating ``clone an administrator's permissions to a guest account'' as a separate behavior, the full pipeline correctly recovers T1098 (Account Manipulation). This example illustrates how behavior decomposition helps recover an individual action embedded in a compound description.

\paragraph{Verification.}
In the Operation Zero Disco report, the proposal stage maps ``Use spoofed IPs and Mac addresses in attacks'' to T1090 (Proxy), but verification rejects this candidate. The evidence does not describe forwarding traffic through a proxy or intermediary, as required by the MITRE ATT\&CK definition of T1090. This example illustrates how verification filters out a candidate that is broadly related to the observed behavior but lacks evidence for the specific technique.

However, verification also falsely rejects 24 gold report--technique pairs in \tool-Bench. For example, ``shellcode is executed via the WinAPI function EnumDeviceMonitor'' (gold: T1055 Process Injection) is rejected because the span alone does not support injection into another process.

\paragraph{Span localization.}
In the Horabot report, a sentence describes requesting a data stream from a C2 server and subsequently exfiltrating encoded email addresses. The verifier in the \textit{w/o Span Localization} ablation rejects T1041 (Exfiltration Over C2 Channel) as the atomic behavior retains only the request to the C2 server, omitting the exfiltration action. The full pipeline accepts T1041 as span localization retrieves the original sentence, restoring the explicit link between C2 communication and data exfiltration. This case illustrates how localized evidence can recover critical details.

\section{Computational Cost}
\label{app:cost}
 
\begin{table*}[t]
\centering
\caption{Per-report inference cost on \tool-Bench with the GPT-4o backbone.}
\label{tab:cost}
\small
\renewcommand\arraystretch{1.1}
\setlength{\tabcolsep}{4pt}
\begin{tabular}{l|cccc}
\Xhline{1.2pt}
\rowcolor{tablerow}
\textbf{Method} & \textbf{LLM calls/report} & \textbf{Tokens (in/out)} & \textbf{Cost/report} & \textbf{Latency} \\
\Xhline{1.2pt}
\rowcolor{gray!8}
One-Shot & 1.2  & 2.3K / 1.2K  & \$0.018 & 7 s \\
CoT      & 1.0  & 1.7K / 0.7K  & \$0.011 & 6 s \\
\rowcolor{gray!8}
\tool    & 48.4 & 61.0K / 7.4K & \$0.226 & 92 s \\
\Xhline{1.2pt}
\end{tabular}
\end{table*}
 
Table~\ref{tab:cost} reports per-report cost with the GPT-4o backbone, averaged over \tool-Bench, at API pricing as of July 2026. \tool~issues 48.4 LLM calls per report, about 20 times the per-report cost of CoT (\$0.226 vs.\ \$0.011) and 15 times its latency (92 s vs.\ 6 s).
 
This overhead matches the intended deployment profile. TTP extraction from published CTI reports is an offline, low-throughput workload: an analyst team processes tens of reports per day, so the total API cost stays within a few dollars, and the per-report latency is minor next to the time an analyst spends reading the report. The error-cost asymmetry also favors accuracy over per-report cost. A hallucinated technique that propagates into a threat-hunting playbook can consume analyst hours, while the entire per-report budget is \$0.23.
\end{document}